\pdfoutput=1

\documentclass[11pt]{article}
\usepackage[table,dvipsnames]{xcolor}
\usepackage{acl}

\usepackage{url}

\usepackage{times}
\usepackage{latexsym}
\usepackage[T1]{fontenc}

\usepackage[utf8]{inputenc}

\usepackage{microtype}

\usepackage{inconsolata}

\usepackage{graphicx}               
\usepackage{tabularx}               
\newcolumntype{C}{>{\centering\arraybackslash}X}
\usepackage{multirow}               
\usepackage{diagbox}                
\usepackage{hhline}                 
\usepackage{color}                  
\usepackage{amsmath,amsthm}                
\usepackage{bm}
\usepackage{amssymb}                
\usepackage{mathtools}              
\usepackage{enumitem}               
\usepackage{booktabs}
\usepackage{subcaption}
\usepackage{bbm}

\makeatletter
\newcommand*\bigcdot{\mathpalette\bigcdot@{.5}}
\newcommand*\bigcdot@[2]{\mathbin{\vcenter{\hbox{\scalebox{#2}{$\m@th#1\bullet$}}}}}
\makeatother

\usepackage{xparse}
\NewDocumentCommand{\heng}{ mO{} }{\textcolor{OrangeRed}{\textsuperscript{\textit{Heng}}\textsf{\textbf{\small[#1]}}}}
\NewDocumentCommand{\qianying}{ mO{} }{\textcolor{CadetBlue}{\textsuperscript{\textit{Qianying}}\textsf{\textbf{\small[#1]}}}}
\NewDocumentCommand{\fei}{ mO{} }{\textcolor{Blue}{\textsuperscript{\textit{Fei}}\textsf{\textbf{\small[#1]}}}}
\NewDocumentCommand{\lingfei}{ mO{} }{\textcolor{Cyan}{\textsuperscript{\textit{Lingfei}}\textsf{\textbf{\small[#1]}}}}

\NewDocumentCommand{\aysa}{ mO{} }{\textcolor{Purple}{\textsuperscript{\textit{Aysa}}\textsf{\textbf{\small[#1]}}}}

 \usepackage{tabularray}
\usepackage{lipsum}  
\usepackage{longtable}
\usepackage{subcaption}
\usepackage{adjustbox}
\usepackage{tikz}
\usepackage{listings}
\usepackage{xcolor,colortbl}

\colorlet{punct}{red!60!black}
\definecolor{background}{HTML}{EEEEEE}
\definecolor{delim}{RGB}{20,105,176}
\definecolor{mycolor}{RGB}{173,216,230}

\colorlet{numb}{magenta!60!black}

\lstdefinelanguage{json}{
    basicstyle=\scriptsize\ttfamily,
    numbers=left,
    numberstyle=\scriptsize,
    stepnumber=1,
    numbersep=8pt,
    showstringspaces=false,
    breaklines=true,
    frame=lines,
    backgroundcolor=\color{background},
    literate=
     *{0}{{{\color{numb}0}}}{1}
      {1}{{{\color{numb}1}}}{1}
      {2}{{{\color{numb}2}}}{1}
      {3}{{{\color{numb}3}}}{1}
      {4}{{{\color{numb}4}}}{1}
      {5}{{{\color{numb}5}}}{1}
      {6}{{{\color{numb}6}}}{1}
      {7}{{{\color{numb}7}}}{1}
      {8}{{{\color{numb}8}}}{1}
      {9}{{{\color{numb}9}}}{1}
      {:}{{{\color{punct}{:}}}}{1}
      {,}{{{\color{punct}{,}}}}{1}
      {\{}{{{\color{delim}{\{}}}}{1}
      {\}}{{{\color{delim}{\}}}}}{1}
      {[}{{{\color{delim}{[}}}}{1}
      {]}{{{\color{delim}{]}}}}{1},
}

\definecolor{c0}{rgb}{0.0500,0.0400,0.0300}
\definecolor{c1}{rgb}{0.0500,0.0400,0.0300}
\definecolor{c2}{rgb}{0.0500,0.0400,0.0300}
\definecolor{c3}{rgb}{0.0500,0.0400,0.0300}
\definecolor{c4}{rgb}{0.0500,0.0400,0.0300}
\definecolor{c5}{rgb}{0.0500,0.0400,0.0300}
\definecolor{c6}{rgb}{0.0500,0.0400,0.0300}
\definecolor{c7}{rgb}{0.0500,0.0400,0.0300}
\definecolor{c8}{rgb}{0.0500,0.0400,0.0300}
\definecolor{c9}{rgb}{0.0500,0.0400,0.0300}
\definecolor{c10}{rgb}{0.1281,0.0867,0.1140}
\definecolor{c11}{rgb}{0.1281,0.0867,0.1140}
\definecolor{c12}{rgb}{0.1281,0.0867,0.1140}
\definecolor{c13}{rgb}{0.1281,0.0867,0.1140}
\definecolor{c14}{rgb}{0.1281,0.0867,0.1140}
\definecolor{c15}{rgb}{0.1281,0.0867,0.1140}
\definecolor{c16}{rgb}{0.1281,0.0867,0.1140}
\definecolor{c17}{rgb}{0.1281,0.0867,0.1140}
\definecolor{c18}{rgb}{0.1281,0.0867,0.1140}
\definecolor{c19}{rgb}{0.1281,0.0867,0.1140}
\definecolor{c20}{rgb}{0.2691,0.2115,0.3736}
\definecolor{c21}{rgb}{0.2691,0.2115,0.3736}
\definecolor{c22}{rgb}{0.2691,0.2115,0.3736}
\definecolor{c23}{rgb}{0.2691,0.2115,0.3736}
\definecolor{c24}{rgb}{0.2691,0.2115,0.3736}
\definecolor{c25}{rgb}{0.2691,0.2115,0.3736}
\definecolor{c26}{rgb}{0.2691,0.2115,0.3736}
\definecolor{c27}{rgb}{0.2691,0.2115,0.3736}
\definecolor{c28}{rgb}{0.2691,0.2115,0.3736}
\definecolor{c29}{rgb}{0.2691,0.2115,0.3736}
\definecolor{c30}{rgb}{0.2897,0.3675,0.6362}
\definecolor{c31}{rgb}{0.2897,0.3675,0.6362}
\definecolor{c32}{rgb}{0.2897,0.3675,0.6362}
\definecolor{c33}{rgb}{0.2897,0.3675,0.6362}
\definecolor{c34}{rgb}{0.2897,0.3675,0.6362}
\definecolor{c35}{rgb}{0.2897,0.3675,0.6362}
\definecolor{c36}{rgb}{0.2897,0.3675,0.6362}
\definecolor{c37}{rgb}{0.2897,0.3675,0.6362}
\definecolor{c38}{rgb}{0.2897,0.3675,0.6362}
\definecolor{c39}{rgb}{0.2897,0.3675,0.6362}
\definecolor{c40}{rgb}{0.2628,0.4655,0.6737}
\definecolor{c41}{rgb}{0.2628,0.4655,0.6737}
\definecolor{c42}{rgb}{0.2628,0.4655,0.6737}
\definecolor{c43}{rgb}{0.2628,0.4655,0.6737}
\definecolor{c44}{rgb}{0.2628,0.4655,0.6737}
\definecolor{c45}{rgb}{0.2628,0.4655,0.6737}
\definecolor{c46}{rgb}{0.2628,0.4655,0.6737}
\definecolor{c47}{rgb}{0.2628,0.4655,0.6737}
\definecolor{c48}{rgb}{0.2628,0.4655,0.6737}
\definecolor{c49}{rgb}{0.2628,0.4655,0.6737}
\definecolor{c50}{rgb}{0.2563,0.5407,0.6901}
\definecolor{c51}{rgb}{0.2563,0.5407,0.6901}
\definecolor{c52}{rgb}{0.2563,0.5407,0.6901}
\definecolor{c53}{rgb}{0.2563,0.5407,0.6901}
\definecolor{c54}{rgb}{0.2563,0.5407,0.6901}
\definecolor{c55}{rgb}{0.2563,0.5407,0.6901}
\definecolor{c56}{rgb}{0.2563,0.5407,0.6901}
\definecolor{c57}{rgb}{0.2563,0.5407,0.6901}
\definecolor{c58}{rgb}{0.2563,0.5407,0.6901}
\definecolor{c59}{rgb}{0.2563,0.5407,0.6901}
\definecolor{c60}{rgb}{0.3706,0.8323,0.7278}
\definecolor{c61}{rgb}{0.3706,0.8323,0.7278}
\definecolor{c62}{rgb}{0.3706,0.8323,0.7278}
\definecolor{c63}{rgb}{0.3706,0.8323,0.7278}
\definecolor{c64}{rgb}{0.3706,0.8323,0.7278}
\definecolor{c65}{rgb}{0.3706,0.8323,0.7278}
\definecolor{c66}{rgb}{0.3706,0.8323,0.7278}
\definecolor{c67}{rgb}{0.3706,0.8323,0.7278}
\definecolor{c68}{rgb}{0.3706,0.8323,0.7278}
\definecolor{c69}{rgb}{0.3706,0.8323,0.7278}
\definecolor{c70}{rgb}{0.4356,0.8630,0.7264}
\definecolor{c71}{rgb}{0.4356,0.8630,0.7264}
\definecolor{c72}{rgb}{0.4356,0.8630,0.7264}
\definecolor{c73}{rgb}{0.4356,0.8630,0.7264}
\definecolor{c74}{rgb}{0.4356,0.8630,0.7264}
\definecolor{c75}{rgb}{0.4356,0.8630,0.7264}
\definecolor{c76}{rgb}{0.4356,0.8630,0.7264}
\definecolor{c77}{rgb}{0.4356,0.8630,0.7264}
\definecolor{c78}{rgb}{0.4356,0.8630,0.7264}
\definecolor{c79}{rgb}{0.4356,0.8630,0.7264}
\definecolor{c80}{rgb}{0.7933,0.9595,0.8424}
\definecolor{c81}{rgb}{0.7933,0.9595,0.8424}
\definecolor{c82}{rgb}{0.7933,0.9595,0.8424}
\definecolor{c83}{rgb}{0.7933,0.9595,0.8424}
\definecolor{c84}{rgb}{0.7933,0.9595,0.8424}
\definecolor{c85}{rgb}{0.7933,0.9595,0.8424}
\definecolor{c86}{rgb}{0.7933,0.9595,0.8424}
\definecolor{c87}{rgb}{0.7933,0.9595,0.8424}
\definecolor{c88}{rgb}{0.7933,0.9595,0.8424}
\definecolor{c89}{rgb}{0.7933,0.9595,0.8424}
\definecolor{c90}{rgb}{0.8906,0.9966,0.9210}
\definecolor{c91}{rgb}{0.8906,0.9966,0.9210}
\definecolor{c92}{rgb}{0.8906,0.9966,0.9210}
\definecolor{c93}{rgb}{0.8906,0.9966,0.9210}
\definecolor{c94}{rgb}{0.8906,0.9966,0.9210}
\definecolor{c95}{rgb}{0.8906,0.9966,0.9210}
\definecolor{c96}{rgb}{0.8906,0.9966,0.9210}
\definecolor{c97}{rgb}{0.8906,0.9966,0.9210}
\definecolor{c98}{rgb}{0.8906,0.9966,0.9210}
\definecolor{c99}{rgb}{0.8906,0.9966,0.9210}
\newcommand{\ourdata}{Ads-XMC\xspace}

\usepackage[framemethod=tikz]{mdframed}

\usepackage{xspace}
\usepackage{comment}
\usepackage{url}

\usepackage[utf8]{inputenc}
\usepackage{algorithm}
\usepackage[commentColor=blue]{algpseudocodex}

\title{From Lazy to Prolific: Tackling Missing Labels in Open Vocabulary Extreme Classification by Positive-Unlabeled Sequence Learning}
\author{Ranran Haoran Zhang$^1$, Bensu Uçar$^2$, Soumik Dey$^2$, Hansi Wu$^2$, Binbin Li$^2$,  Rui Zhang$^1$ \\
$^1$ Penn State University \\
$^2$ eBay Inc\\
}
\date{}

\begin{document}

\maketitle



\begin{abstract}

Open-vocabulary Extreme Multi-label Classification (OXMC) extends traditional XMC by allowing prediction beyond an extremely large, predefined label set (typically $10^3$ to $10^{12}$ labels), addressing the dynamic nature of real-world labeling tasks. However, self-selection bias in data annotation leads to significant missing labels in both training and test data, particularly for less popular inputs. This creates two critical challenges: generation models learn to be ``lazy'' by under-generating labels, and evaluation becomes unreliable due to insufficient annotation in the test set. In this work, we introduce Positive-Unlabeled Sequence Learning (PUSL), which reframes OXMC as an infinite keyphrase generation task, addressing the generation model's laziness. Additionally, we propose to adopt a suite of evaluation metrics, F1@$\mathcal{O}$ and newly proposed B@$k$, to reliably assess OXMC models with incomplete ground truths. In a highly imbalanced e-commerce dataset with substantial missing labels, PUSL generates 30\% more unique labels, and 72\% of its predictions align with actual user queries. On the less skewed EURLex-4.3k dataset, PUSL demonstrates superior F1 scores, especially as label counts increase from 15 to 30. Our approach effectively tackles both the modeling and evaluation challenges in OXMC with missing labels. Our data split was released in huggingface datasets\footnote{\url{https://huggingface.co/datasets/windchimeran/pusl}}

\end{abstract}

\section{Introduction}

\begin{figure}[t!]
  \centering
    \includegraphics[width=0.8\linewidth]{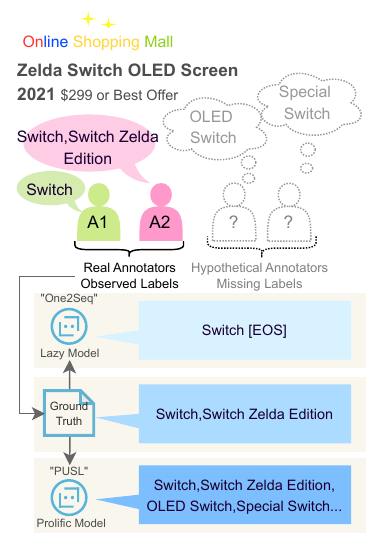}

\caption{Real annotators often provide limited labels (e.g., Switch, Switch Zelda Edition), while many potential labels remain uncaptured (e.g., OLED Switch, Special Switch). This gap between observed and expected labels misleads generation models to be lazy by prematurely terminating label generation. 
Our proposed PUSL resolves the model laziness problem and can learn from incomplete ground truth. 
}
\label{fig:pusl_intro}
\vspace{-6mm}
\end{figure}

Extreme multi-label classification (XMC) is a challenging and critical task in natural language processing, with wide-ranging applications such as e-commerce query recommendations \cite{jain2019slice,zhou2019domain,song2021triangular,zhang2023semantic, ashirbad-etal-2024}, tag recommendation for Wikipedia articles \cite{pmlr-v9-dekel10a}, legal document \cite{chalkidis-etal-2019-large}, stackoverflow questions \cite{kang2021leveraging} and social media posts \cite{wang2019topic,yu2023generating}. 
XMC aims to assign appropriate labels to each input from a vast pool of possibilities, often numbering from thousands to trillions.

Open-vocabulary Extreme Multi-label Classification (OXMC) bridges the traditional XMC and well-studied keyphrase generation by introducing the generation models \cite{bhatia2015sparse, zhang2016keyphrase,prabhu2018extreme,wang2019topic,chang2021extreme} into the XMC framework \cite{simig-etal-2022-open}. While XMC assigns labels from a vast, predefined set, OXMC recognizes that label sets are continuously evolving, with new labels emerging over time. This shift acknowledges that predefined label sets, no matter how extensive, may not capture all possible ways to categorize input instances, especially in rapidly changing domains such as e-commerce and social media. Due to the larger scale of the datasets, successfully addressing the challenges posed by OXMC could also benefit the general keyphrase generation area. 

However, our research reveals that the prevalence of missing labels in OXMC training data leads to ``lazy'' generation models which predict only a few labels per input. This problem stems from the self-selection bias in the data curation processes of OXMC, as illustrated in Figure~\ref{fig:pusl_intro}. Annotators tend to selectively tag data points, with popular inputs receiving more annotations. This self-selection bias causes insufficient annotators and labels on most data. In our dataset (\ourdata), 89\% of data points have fewer than five labels, with only 15\% of these ever having sufficient annotators. In training, these cause the models to prematurely stop label generation before producing a sufficient number of labels. Moreover, existing evaluation metrics fail to accurately reflect this problem in test data, further exacerbating the issue.

To address the generation model's ``laziness'' problem, we introduce Positive-Unlabeled Sequence Learning (PUSL) to foster prolific and accurate label prediction. PUSL treats observed labels as positive, and treats potential but unobserved labels as unlabeled, focusing on positive data to infer a hypothetically infinite sequence of labels. This method views existing labels as part of a larger conceivable set, aiming to mirror exhaustive annotations and mitigate bias from training data. By expanding beyond initially observed labels, PUSL overcomes gaps from insufficient annotations, increasing the diversity of labels generated across all inputs.
To further enhance PUSL's capability in generating diverse labels, we implement a post-training phase that leverages both original diverse training data and augmented data created through rejection sampling, exposing the model to a more comprehensive label set. 

To address the challenges in evaluating OXMC models with incomplete ground truth due to missing annotations, we propose two metrics: F1@$\mathcal{O}$ and B@$k$. Traditional metrics like F1@k tend to favor lazy models when faced with incomplete test data. Our proposed F1@$\mathcal{O}$ mitigates this bias by adapting to the variable number of available ground truth labels, avoiding penalties for prolific models. Additionally, B@$k$ addresses the limitations of existing metrics by penalizing under-generation while fairly assessing predictions beyond available ground truth. Together, these metrics provide a more faithful assessment of OXMC models in real-world scenarios with incomplete annotations, effectively distinguishing between lazy and prolific prediction behaviors.

Our experimental results demonstrate PUSL's effectiveness and applicability across various domains. In the highly-imbalanced \ourdata dataset with substantial missing labels, PUSL generates 30\% more unique labels per item, and 72\% of its predicted keyphrases align with actual user queries. In the EURLex4.3k dataset, with an average of 15 keyphrases per data point, PUSL shows promising results over all baselines in F1@$\mathcal{O}$ and B@$k$, particularly as $k$ increases from 15 to 30. Our contributions include:
\begin{itemize}
\setlength{\itemsep}{-0.5em}

\item \textbf{Data Imbalance Analysis:} We identify self-selection bias in OXMC datasets as the cause of substantial missing labels.

\item \textbf{Model Solution:} We recast OXMC as Positive Unlabeled Sequence Learning, and our experiment demonstrates its effectiveness to foster prolific and accurate label prediction.

\item \textbf{Evaluation Solution:} We propose F1@$\mathcal{O}$ and B@$k$ metrics to reliably evaluate models despite incomplete ground-truth labels.
\end{itemize}

\begin{figure*}[t!]
  \centering
  \begin{minipage}[b]{0.45\textwidth}
  \centering
      \begin{subfigure}[b]{0.75\linewidth}
        \centering
        \includegraphics[width=\linewidth]{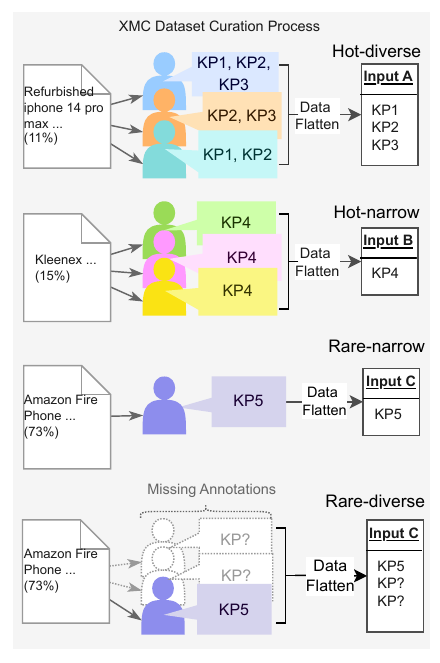}
        \caption{The XMC dataset curation process. KP stand for keyphrase. Popular inputs like iPhone or Kleenex receive more annotations, while others are neglected.}
        \label{fig:budgetatk}
      \end{subfigure}
  \end{minipage}
  \hfill
\begin{minipage}[b]{0.45\textwidth}

      \begin{subfigure}[b]{\linewidth}
        \centering
        \includegraphics[width=\linewidth]{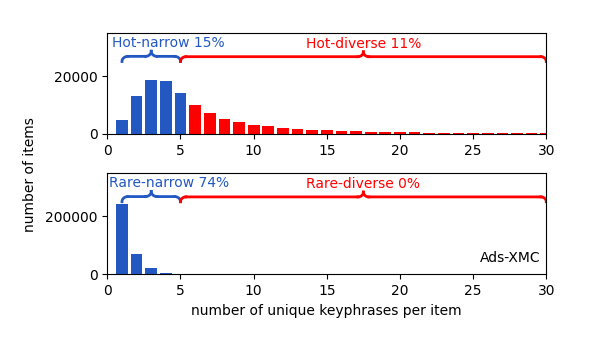}
        \caption{Distribution of numbers of items by unique keyphrases in Ads-XMC.}
        \label{fig:hot_rare}
      \end{subfigure}
          
      \begin{subfigure}[b]{\linewidth}
        \centering
        \includegraphics[width=\linewidth]{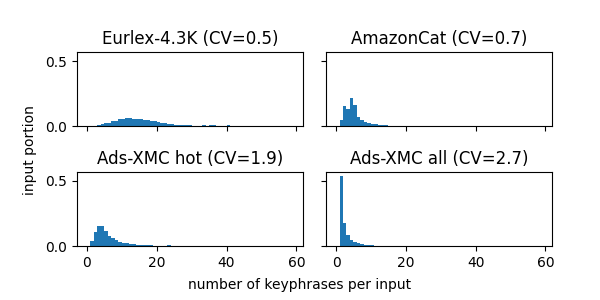}
        \caption{Frequency distribution of the number of keyphrases per input for various XMC datasets.  Higher CVs indicating greater disparity in the number of keyphrases per input.}
        \label{fig:other_data_dist}
      \end{subfigure}
\end{minipage}
      \caption{Analysis of Missing Labels in OXMC.}
\vspace{-3mm} 
\label{fig:analysis_xmc}
\end{figure*}

\section{Related Work} 
\paragraph{Missing Labels in XMC.}
The large label set in XMC challenges exhaustive annotation, resulting in two types of missing labels: insufficient labels in the training set \cite{schultheis2021unbiased,schultheis2022missing,schultheis2023consistent} and unseen labels in the test set \cite{gupta2021generalized}. Existing approaches like GROOV \cite{simig-etal-2022-open} address unseen labels but tend to predict fewer, while XLGen \cite{jung-etal-2023-cluster} is limited to seen labels due to its pre-clustering process.
Previous studies attribute missing labels to annotator subjectivity and heedlessness \cite{sterckx2016supervised,lei2021keyphrase}. \citet{wei2019does} theoretically analyze the impact of missing labels under the Missing Completely at Random (MCAR) assumption \cite{rubin1976inference}.
By contrast, our investigation reveals that the missing label mechanism in XMC deviates from the commonly assumed MCAR, because it stems from the unequal distribution of annotations from self-selection bias \cite{marlin2012collaborative}. We introduce an improved generation model capable of addressing both unseen and insufficient labels by explicitly modeling this distribution, empowering XMC models for the complexities of real-world data.



\paragraph{Positive Unlabeled Learning.}

Traditional PU learning trains a binary classifier to differentiate between positive and negative examples without labeled negatives in the dataset \cite{lee2003learning,liu2003building,elkan2008learning,kiryo2017positive}. This has evolved into Multi-Positive and Unlabeled learning for multi-class scenarios, where training data includes labeled examples from several positive classes and unlabeled examples which may be positive or a single negative class \cite{xu2017multi}.
In XMC, the large label space precludes simple conversion to multiple binary classifications, while our proposed PUSL leverages open vocabulary generation models to make this task feasible. Moreover, our investigation reveals that the missing label mechanism in XMC deviates from the commonly assumed MCAR condition in PU learning, aligning more with Selected At Random PU Learning \cite{bekker2019beyond,bekker2020learning}, indicating systematic patterns or biases in label absence.

\section{Analysis on Missing Label in OXMC}
\label{sec:pilot}
To investigate the effect of self-selection bias on the label distribution, we analyzed an e-commerce dataset, \ourdata. Unlike most public OXMC datasets, \ourdata retains crucial information on user interaction frequencies (i.e., annotation frequency for each input), making it uniquely suitable for this analysis. The dataset consists of item-keyphrase pairs derived from user interactions, along with interaction frequencies. During the OXMC data curation process, as shown in Figure~\ref{fig:budgetatk}, the raw data is transformed into a flattened dataset by deduplicating labels and retaining only a single instance of each unique label per item, aligning with the format used in previous public datasets. Thus, the number of keyphrases is bounded by the total user interaction, if no duplication. 

We classified data based on item popularity (hot or rare) and label diversity (diverse or narrow) using a threshold of five interactions or labels, identified as a significant breakpoint for user interest. Hot items have $\geq$ 5 interactions (top 26\%), while diverse items have $\geq$ 5 unique labels. This categorization data samples in four groups: (1) \textbf{Hot-diverse}: High interaction with a wide variety of labels; (2) \textbf{Hot-narrow}: Many interactions but a limited range of labels; (3) \textbf{Rare-narrow}: Few interactions and a small set of labels; and (4) \textbf{Rare-diverse}: Potentially broad label spectrum if more interactions occurred.

As shown in Figure~\ref{fig:hot_rare}, Rare-narrow items cover 73.7\% of the dataset, while Hot-narrow items, despite having few unique keyphrases per item, cover only 15.2\%. Annotators' personal choices lead to an uneven distribution of attention, similar to the 80/20 rule: a small number of widely recognized items attract most focus, while the majority remain less popular and under-annotated. This results in missing labels, affecting both training and test data in OXMC datasets, and posing significant challenges for model training and evaluation.

The issue of \emph{self-selection bias} (unequal distribution of annotations) is pervasive in OXMC datasets, though most lack the frequency data to quantify this imbalance. For example, the \textit{Barack Obama} Wikipedia page likely receives more revisions and annotations than the less-known \textit{Andrew Dunlop} page, resulting in a more comprehensive tag set for the former. Similarly, on music streaming platforms, tracks by well-known artists like \textit{Beyoncé} are tagged with more genres and moods than those by lesser-known musicians. 

Although there is no annotation frequency information in other datasets, we quantify the label missingness by the coefficient of variation (CV) \cite{Brown1998}: $CV = \mu/ \sigma$, where $\mu$ is the mean number of keyphrases and $\sigma$ is the standard deviation. A higher CV indicates more imbalance. Figure~\ref{fig:other_data_dist} shows that OXMC datasets have different levels of keyphrase imbalance, with \ourdata being the most imbalanced. The \ourdata hot split has a distribution similar to AmazonCat and a lower CV than the \ourdata all split, demonstrating different degrees of missing labels.
Key takeaways from the analysis: 
\begin{itemize}
\setlength{\itemsep}{-0.5em}
    \item Missing labels result from self-selection bias, with few items receiving most labels.
    \item Both training and test data suffer from missing labels, making model training and evaluation challenging.
    \item The imbalance and severity of missing labels varies across datasets, measurable by the coefficient of variation of keyphrases per input.
\end{itemize}

\section{Method}
We present our PUSL approach to address the challenges of missing labels in OXMC. Additionally, we propose evaluation metrics that account for missing labels in the test data, enabling a more accurate assessment of model performance.

\subsection{Task: OXMC as Positive-Unlabeled Sequence Generation}

Given a textual input $x$, XMC aims to predict a set of labels $\mathcal{Y} = \{y_1, y_2, \ldots, y_{|\mathcal{Y}|}\}$ from an extremely large label space. Each label $y_j = [w_1, w_2, \ldots, w_{|y_j|}]$ is a sequence of words (i.e., keyphrase). We formulate OXMC as a Positive-Unlabeled Sequence Generation problem:
\begin{equation}
\begin{split}
    \mathcal{D} &= \{(x_i, \mathcal{Y}^p_i)\}_{i=1}^n  \bigcup \{(x_i, \mathcal{Y}^u_i)\}_{i=1}^n \\
    \mathcal{D}_{train} &\subset \{(x_i, \mathcal{Y}^p_i)\}_{i=1}^n,     \mathcal{D}_{test} \subset \{(x_i, \mathcal{Y}^p_i)\}_{i=1}^n 
     \end{split}
\end{equation}
where $\mathcal{Y}^p_i$ is the observed label set and $\mathcal{Y}^u_i$ is the unobserved label set for document $x_i$. 

Real-world XMC datasets typically only include observed keyphrases due to annotation challenges. The goal is to learn a top-$k$ keyphrase generator $f: (x,k) \rightarrow \mathcal{Y}$ that maps $x$ to $\{y_1, y_2, \ldots, y_k \}$, where $k$ is a predefined requirement (e.g., determined by the constraints of the user interface or website), but may exceed the number of observed keyphrases in most labeled data.

\subsection{Background: Keyphrase Generation Biases}
\begin{figure}[t!]
  \centering
    \includegraphics[width=0.95\linewidth]{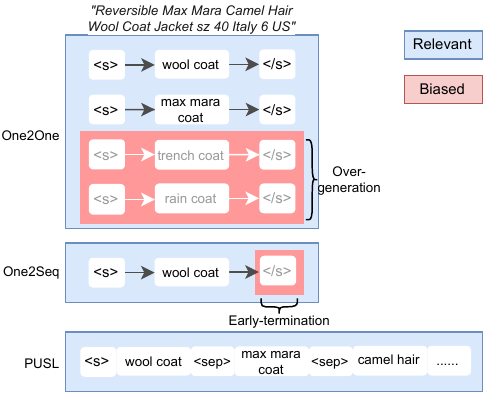}
    \caption{Comparison of biases in keyphrase generation models: One2Seq's early termination, One2One's over-generation, and PUSL.
    }
    \label{fig:three_paradigms} 
    \vspace{-4mm}
\end{figure}

Existing keyphrase generation paradigms face biases when applied to OXMC tasks due to incomplete training data. We discuss two main paradigms and their associated biases as shown in Figure~\ref{fig:three_paradigms}:
\paragraph{One2Seq.} This paradigm treats keyphrase generation as a single sequence generation task~\cite{chan2019neural,yuan-etal-2020-one}. The model learns to generate all keyphrases as a concatenated sequence, separated by delimiters, and terminated by an end-of-sequence (EOS) token. \textbf{GROOV}, the state-of-the-art generation One2Seq model for OXMC, employs a label trie and multi-softmax loss for unordered training~\cite{simig-etal-2022-open}. 
\textbf{Early Termination Bias.} When trained on incomplete XMC data, One2Seq models tend to prematurely generate the EOS token, resulting in shorter keyphrase sequences. This bias stems from the model learning to terminate generation based on the limited number of observed keyphrases in the training data, rather than the true, exhaustive set of relevant keyphrases.

\paragraph{One2One.} In this paradigm, each training instance consists of an input document paired with only one keyphrase~\cite{meng-etal-2017-deep,chen-etal-2019-integrated,Chen_Gao_Zhang_King_Lyu_2019}. During inference, the model generates top-$k$ candidate keyphrases through beam search.
\textbf{Over-generation Bias.} 
One2One models often exhibit over-generation bias, especially when producing more keyphrases than the input document can support. This bias arises from the beam search decoding process \cite{kang2021leveraging}, which can generate keyphrases without considering their absolute probabilities, potentially including low-probability, irrelevant candidates. When the requested number of keyphrases exceeds the relevant ones, the model may produce low-quality or irrelevant keyphrases to meet the quota, leading to their inclusion in the final output.

\subsection{Model Solution: Positive-Unlabeled Sequence Learning}
\paragraph{Training.}
To tackle the missing label problem, our PUSL model treats the training target keyphrases as an infinite sequence. Unlike standard sequence-to-sequence models that use an end-of-sequence (EOS) token, PUSL omits this token, allowing the model to generate keyphrases continuously without a defined endpoint.

\paragraph{Post Training. }
PUSL exhibits a repetition problem when generating keyphrases beyond the input's natural limit, similar to cognitive overload. This issue occurs when the model is pushed to produce more keyphrases than the input supports. To address its repetition problem and enhance the keyphrase generation, we implement a post-training phase using two data sources: (1) Original Train-Diverse data: We select instances from the original training set with an above-average number of keyphrases, focusing on richer annotations. (2) Augmented data from rejection sampling: We use PUSL to generate additional keyphrases for instances with few keyphrases, then apply rejection sampling to select those where the total number (original + generated) exceeds the dataset mean after deduplication. This creates a synthetic dataset with a higher density of plausible keyphrases.

\paragraph{Inference.} PUSL employs an auto-regressive decoding strategy for keyphrase generation. Although designed for infinite generation, we extract only the top-$k$ results by adjusting the logits processor during decoding. Algorithm~\ref{algo:pusl_decoding} outlines this process, which uses Begin-of-Keyphrase (BOK) and End-of-Keyphrase (EOK) tokens to delineate boundaries. This approach, as opposed to using a single separator (SEP), simplifies the decoding logit processor implementation and reduces the occurrence of empty keyphrases.


\begin{algorithm}[t!]
\small
\caption{Top-$k$ PUSL Decoding}
\begin{algorithmic}[1]
\Require $\text{token}_{1:n-1}$, $\text{logits}$, k
\State $\text{eok\_count} \gets \text{count}(\text{tokens}_{1:n-1}, \text{EOK})$
\State $\text{completed\_sequences} \gets \text{eos\_counts} \geq k$

\State $\text{completed\_keyphrase} \gets \text{token}_{n-1} == \text{EOK}$
\LComment{Force the model to decode a new keyphrase after generating $\text{EOK}$}
\State $\text{logits[completed\_keyphrase, :]} \gets -\infty$
\State $\text{logits[completed\_keyphrase, EOK]} \gets 0$

\LComment{Terminate generation after generating $k$ $\text{EOK}$}

\State $\text{logits[completed\_sequences, :]} \gets -\infty$
\State $\text{logits[completed\_sequences, EOS]} \gets 0$




\State \Return $\text{logits}$
\end{algorithmic}
\label{algo:pusl_decoding}
\end{algorithm}


\subsection{Evaluation Solution: Faithful Metrics for Incomplete Ground Truth}
\label{sec:metrics}
Evaluating OXMC models is challenging due to missing labels in the test data, which contains data points with fewer keyphrases than expected. We begin by discussing the limitations of existing metrics in the presence of missing labels and then propose new metrics for more accurate comparison.

\begin{figure}[t!]
  \centering
    \includegraphics[width=0.85\linewidth]{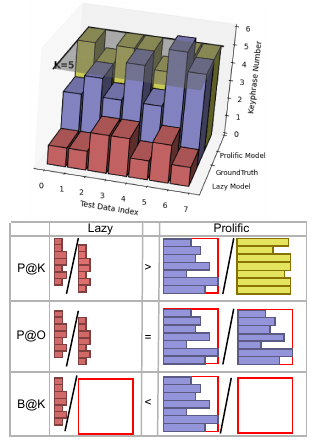}

\caption{Comparison of evaluation metrics for lazy and prolific models under incomplete ground truth. Ground truth beyond K are truncated in the denominator. Traditional P@K favors lazy models even when all predictions are relevant. Proposed P@O equalizes performance between lazy and prolific models, while B@K penalizes under-generation. These metrics provide a more faithful evaluation with missing ground truth. 
}
  \label{fig:metrics} 
  \vspace{-4mm}
\end{figure}

Traditional metrics like Precision@k (P@k) use a fixed k across all test data points, which can unfairly favor ``lazy'' models that generate fewer keyphrases, especially in datasets with many missing labels. As shown in Figure~\ref{fig:metrics}, even if both lazy and prolific models are perfect predictors (only generating keyphrases in the ground truth), the lazy model achieves higher precision scores due to smaller denominators, while the prolific model is penalized for generating more keyphrases than are present in the ground truth.

To address this issue, we adopt F1@$\mathcal{O}$~\cite{yuan-etal-2020-one} which is the F1 score of the following:\footnote{Precion@$\mathcal{O}$ is identical to R-Precision@k \cite{schutze2008introduction,chalkidis-etal-2021-multieurlex}}
\begin{equation}
\begin{split}
P@\mathcal{O} &= \frac{|\hat{\mathcal{Y}}_{:\mathcal{O}} \cap \mathcal{Y}|}{|\hat{\mathcal{Y}}_{:\mathcal{O}}|} \quad R@\mathcal{O} = \frac{|\hat{\mathcal{Y}}_{:\mathcal{O}} \cap \mathcal{Y}|}{|\mathcal{Y}|}
\end{split}
\end{equation}
where $|\cdot|$ denotes the cardinality of a set, $\mathcal{Y}$ is the ground truth,
$\mathcal{O}$ is the number ground truth labels, $\hat{\mathcal{Y}}_{:\mathcal{O}}$ is model's top-$\mathcal{O}$ predicted keyphrases.
As shown in Figure~\ref{fig:metrics}, F1@$\mathcal{O}$ adjusts to the varying number of ground truth labels per data point, equalizing the performance of perfect lazy and prolific models. 

\setlength{\tabcolsep}{4pt}
\begin{table*}[!t]
\centering
\small

\begin{tabular}{@{}l|l|rrrrr|rrrrr@{}}
\toprule
Dataset                     & Model      & \multicolumn{5}{c|}{Test-Narrow ($|\mathcal{Y}| \leq 2\mu$)}                                                                                               & \multicolumn{5}{c}{Test-Diverse ($|\mathcal{Y}| > 2\mu$)}                                                                                               \\ \midrule
                            &  & \#K@$\mu$      & \multicolumn{1}{r|}{\#K@$2\mu$}     & B@$\mu$        & \multicolumn{1}{r|}{B@$2\mu$}       & F1@$\mathcal{O}$ & \#K@$\mu$      & \multicolumn{1}{r|}{\#K@$2\mu$}     & B@$\mu$        & \multicolumn{1}{r|}{B@$2\mu$}       & F1@$\mathcal{O}$ \\ \midrule
\multirow{1}{*}{EURLex4.3k }                             & One2One       & $\mu$ & \multicolumn{1}{r|}{$2\mu$}          & \cellcolor{c35} \color{white} 35.15& \multicolumn{1}{r|}{\cellcolor{c22} \color{white} 22.73} & \cellcolor{c36} \color{white} 36.68& $\mu$ & \multicolumn{1}{r|}{$2\mu$}          & \cellcolor{c25} \color{white} 25.98& \multicolumn{1}{r|}{\cellcolor{c15} \color{white} 16.63} & \cellcolor{c15} \color{white}  14.23\\ 

& GROOV      & 7.31           & \multicolumn{1}{r|}{7.56}           & \cellcolor{c24} \color{white} 24.09          & \multicolumn{1}{r|}{\cellcolor{c15} \color{white} 12.14}          & \cellcolor{c36} \color{white} 36.94            &  13.57          & \multicolumn{1}{r|}{17.33}          & \cellcolor{c69}  69.06          & \multicolumn{1}{r|}{\cellcolor{c41} \color{white} 41.42}          & \cellcolor{c43} \color{white} 43.06            \\
\multirow{2}{*}{$(\mu = 15)$ }  & One2Seq    & 14.05          & \multicolumn{1}{r|}{\textbf{22.36}} & \cellcolor{c52} \color{white} 52.01          & \multicolumn{1}{r|}{\cellcolor{c33} \color{white} 33.32}          & \cellcolor{c57} \color{white} 57.77            & 14.88          & \multicolumn{1}{r|}{\textbf{28.98}} & \cellcolor{c77} \color{black} 77.05          & \multicolumn{1}{r|}{\cellcolor{c80} \color{black} 80.02}          & \cellcolor{c74} \color{black} 74.40            \\
                            & PUSL         & \textbf{14.56} & \multicolumn{1}{r|}{22.02}          & \cellcolor{c55} \color{white} 54.75 & \multicolumn{1}{r|}{\cellcolor{c34} \color{white} 33.64} & \cellcolor{c59} \color{white} 58.81   & \textbf{14.95} & \multicolumn{1}{r|}{28.78}          & \cellcolor{c80} \color{black} 88.56 & \multicolumn{1}{r|}{\cellcolor{c82} \color{black} 82.29} & \cellcolor{c80} \color{black} 80.28   \\ \midrule
\multirow{1}{*}{AmazonCat}   & One2One       & $\mu$ & \multicolumn{1}{r|}{$2\mu$}          & \cellcolor{c48} \color{white} 48.37 & \multicolumn{1}{r|}{\cellcolor{c38} \color{white} 38.58} & \cellcolor{c48} \color{white} 48.06 & $\mu$ & \multicolumn{1}{r|}{$2\mu$}          & \cellcolor{c71} \color{black} 70.90 & \multicolumn{1}{r|}{\cellcolor{c60} \color{black} 60.26} & \cellcolor{c52} \color{white} 52.64\\ 
                            & GROOV        & 2.97           & \multicolumn{1}{r|}{2.99}           & \cellcolor{c48} \color{white} 48.33 & \multicolumn{1}{r|}{\cellcolor{c24} \color{white} 24.34}          & \cellcolor{c63} \color{black} 62.54   & 3.58           & \multicolumn{1}{r|}{3.62}           & \cellcolor{c64} \color{black} 63.79          & \multicolumn{1}{r|}{\cellcolor{c32} \color{white} 32.26}          & \cellcolor{c36} \color{white} 35.92            \\
\multirow{2}{*}{$(\mu = 5)$ } & One2Seq    & 4.71           & \multicolumn{1}{r|}{8.83}           & \cellcolor{c39} \color{white} 38.74          & \multicolumn{1}{r|}{\cellcolor{c31} \color{white} 30.63}          & \cellcolor{c37} \color{white} 37.50            & 4.97           & \multicolumn{1}{r|}{9.90}           & \cellcolor{c64} \color{black} 64.28          & \multicolumn{1}{r|}{\cellcolor{c58} \color{white} 57.57}          & \cellcolor{c55} \color{white} 55.58            \\
                            & PUSL         & \textbf{4.91}  & \multicolumn{1}{r|}{\textbf{9.49}}  & \cellcolor{c48} \color{white} 47.67          & \multicolumn{1}{r|}{\cellcolor{c39} \color{white} 38.71} & \cellcolor{c47} \color{white} 47.12            & \textbf{5.00}  & \multicolumn{1}{r|}{\textbf{9.97}}  & \cellcolor{c75} \color{black} 74.82 & \multicolumn{1}{r|}{\cellcolor{c69} \color{black} 69.16} & \cellcolor{c61} \color{black} 61.64   \\ \midrule
\multirow{1}{*}{\ourdata}    & One2One       & $\mu$ & \multicolumn{1}{r|}{$2\mu$}          & \cellcolor{c35} \color{white} 34.78 & \multicolumn{1}{r|}{\cellcolor{c24} \color{white} 24.47} & \cellcolor{c35} \color{white} 35.15 & $\mu$ & \multicolumn{1}{r|}{$2\mu$}          & \cellcolor{c59} \color{white} 59.32 & \multicolumn{1}{r|}{\cellcolor{c48} \color{white} 48.27} & \cellcolor{c34} \color{white} 34.37\\ 
                            & GROOV        & 1.03           & \multicolumn{1}{r|}{1.03}           & \cellcolor{c19} \color{white} 18.72          & \multicolumn{1}{r|}{\cellcolor{c9} \color{white} 9.36}           & \cellcolor{c29} \color{white} 28.95            & 1.07           & \multicolumn{1}{r|}{1.07}           & \cellcolor{c26} \color{white} 26.01          & \multicolumn{1}{r|}{\cellcolor{c13} \color{white} 13.00}          & \cellcolor{c13} \color{white} 12.56            \\
\multirow{2}{*}{$(\mu = 3)$ } & One2Seq    & 1.89           & \multicolumn{1}{r|}{2.17}           & \cellcolor{c26} \color{white} 25.52          & \multicolumn{1}{r|}{\cellcolor{c14} \color{white} 13.75}          & \cellcolor{c31} \color{white} 31.26            & 1.85           & \multicolumn{1}{r|}{2.10}           & \cellcolor{c41} \color{white} 41.43          & \multicolumn{1}{r|}{\cellcolor{c23} \color{white} 22.73}          & \cellcolor{c20} \color{white} 20.44            \\
                            & PUSL         & 2.46           & \multicolumn{1}{r|}{3.31}           & \cellcolor{c35} \color{white} 34.50 & \multicolumn{1}{r|}{\cellcolor{c19} \color{white} 19.48} & \cellcolor{c38} \color{white} 38.54   & 2.46           & \multicolumn{1}{r|}{3.23}           & \cellcolor{c56} \color{white} 56.14 & \multicolumn{1}{r|}{\cellcolor{c34} \color{white} 33.79} & \cellcolor{c27} \color{white} 27.94   \\
                            & PUSL+PT        & \textbf{2.78}& \multicolumn{1}{r|}{\textbf{4.27}}  & \cellcolor{c32} \color{white} 34.73& \multicolumn{1}{r|}{\cellcolor{c21} \color{white} 21.03} & \cellcolor{c36} \color{white} 36.47& \textbf{2.76}& \multicolumn{1}{r|}{\textbf{4.17}}  & \cellcolor{c62} \color{black} 62.52& \multicolumn{1}{r|}{\cellcolor{c41} \color{white} 41.21} & \cellcolor{c30} \color{white} 30.09\\ 

                            \bottomrule
\end{tabular}
\caption{Evaluation results for various OXMC datasets. We present F1@$\mathcal{O}$, B@$k$, and number of unique keyphrases (\#K@$k$) where $k$ is chosen as $\mu$ or $2\mu$. Lighter color means higher results. Among all One2Seq family models, PUSL demonstrates the highest prolificacy across all datasets. 
One2One shows high performance in \ourdata, where the keyphrase set is typically narrow but performs poorly on the diverse EURLex4.3k dataset. 
}

\label{tab:gtdependent}
\vspace{-4mm}
\end{table*}

Furthermore, in practice, F1@$\mathcal{O}$ may still favor imperfect lazy models, as predicting more relevant keyphrases is inherently more challenging, and prolific models may be unfairly penalized by missing ground truth.
To provide a more faithful evaluation that considers both fairness and the user-defined $k$, we propose a new metric BudgetAccuracy@$k$ (B@$k$):
\begin{equation}
\text{B}@k = \frac{|\hat{\mathcal{Y}}_{:k} \cap \mathcal{Y}|}{k}
\end{equation} where $\hat{\mathcal{Y}}_{:k}$ represents the model's top-$k$ predicted keyphrases and $\mathcal{Y}$ is the ground truth. B@$k$ addresses variable ground truth sizes and under-generation by models: (1) It penalizes models for generating fewer than $k$ keyphrases by treating missing predictions as empty. (2) It avoids penalizing models for generating more keyphrases than available in the ground truth.
B@$k$ measures the model's ability to generate the desired number of keyphrases while maintaining relevance to the ground truth. The upper bound of B@$k$ is determined by $\min(|\mathcal{Y}|, k) / k$, reflecting the limitations of the available ground truth. This makes B@$k$ particularly suitable for evaluating OXMC tasks where the ground truth may be incomplete.

This suite of metrics addresses challenges posed by incomplete ground truth and varying model behaviors in OXMC tasks. By using these metrics in combination, they enable a more faithful comparison between lazy and prolific models, accurately reflecting a model's ability to generate comprehensive and relevant keyphrase sets.

\section{Experimental Settings}

\subsection{Datasets and Diverse / Narrow Splits}
We evaluate our approach on two public datasets commonly used in extreme multi-label classification research: \textbf{Eurlex 4.3k} \cite{chalkidis-etal-2019-large} and \textbf{AmazonCat} \cite{mcauley2013hidden}. We also include \textbf{\ourdata} derived from search logs, where buyers act as annotators by clicking on items in the search results page for a given query.

Traditional XMC datasets were split to ensure each label appeared at least once in training \cite{gupta2021generalized}, rather than maintaining independent and identically distributed (IID) samples. As OXMC models now handle unseen labels \cite{simig-etal-2022-open}, we introduce a new splitting procedure that prioritizes IID evaluation while enabling analysis across different keyphrase densities. Our procedure (Algorithm~\ref{algo:dataset_split}) re-splits the data into \textbf{Test-Narrow} ($|\mathcal{Y}| \leq 2\mu$) and \textbf{Test-Diverse} ($|\mathcal{Y}| > 2\mu$) based on the mean number of keyphrases $\mu$, allowing systematic evaluation of model performance on inputs with both few and many associated keyphrases. Dataset statistics and detailed splitting process are provided in Appendix~\ref{sec:datadetail}.





\subsection{Baselines and Implementation Details}
We compare PUSL with three state-of-the-art keyphrase generation baselines: (1) self-terminated generation models: \textbf{One2Seq}~\cite{chan2019neural,yuan-etal-2020-one} and \textbf{GROOV}~\cite{simig-etal-2022-open}, and (2) top-$k$ generation models: \textbf{One2One}~\cite{meng-etal-2017-deep,chen-etal-2019-integrated,Chen_Gao_Zhang_King_Lyu_2019}. 
We focus on generation models since traditional classification models cannot predict beyond a predefined label set \cite{simig-etal-2022-open}. 
For fair comparisons, all models use the same t5-base architecture \cite{raffel2020exploring} and identical hyperparameters. We only apply post-training on PUSL for \ourdata dataset, due to the high repetition rate. 
Hyperparameters and post-training ablation studies are provided in Appendix~\ref{appendix:hyper} and \ref{appendix:ppusl}.

\section{Results and Analysis}
We compare PUSL with state-of-the-art baselines across various datasets and metrics, focusing on incomplete ground truth (Section~\ref{sec:exp_intro}), early-termination bias in self-terminated models (Section~\ref{sec:one2seq}), and over-generation bias in top-$k$ models (Section~\ref{sec:one2one}). Our analysis includes standard metrics, human evaluations, and an exhaustive annotation experiment (Section~\ref{sec:exhaustive}), providing comprehensive insights into model performance and dataset characteristics.






\subsection{Evaluation using Incomplete Ground Truth}
\label{sec:exp_intro}
Table~\ref{tab:gtdependent} presents our evaluation results across different datasets and models, highlighting the challenges of evaluation with incomplete ground truth. 

\paragraph{Evaluate with More Ground Truth: Test-Diverse vs Test-Narrow.} Our results show a significant performance gap between the Test-Diverse and Test-Narrow subsets across all models. Test-Diverse, with more ground truth labels, generally yields higher BudgetAccuracy and F1 scores, indicating sufficient annotation. In contrast, Test-Narrow likely suffers from missing ground truth, highlighting the impact of incomplete labels on evaluation metrics.

\paragraph{Ask Models to be More Prolific: B@$\mu$ vs B@2$\mu$.} Comparing B@$\mu$ and B@2$\mu$ evaluates models' ability to generate relevant keyphrases beyond the training data average. In Test-Diverse, prolific models show minimal performance decline from B@$\mu$ to B@2$\mu$, while Test-Narrow exhibits sharp drops due to limited ground truth. The performance gap between B@$\mu$ and B@2$\mu$ also reflects ground truth distribution: a significant drop in \ourdata Test-Narrow versus a smaller decline in AmazonCat Test-Narrow suggests a smoother distribution in the latter, as seen in Figure~\ref{fig:other_data_dist}.

\begin{table}[!t]
\scriptsize
\centering
\begin{tabular}{l|cc|cc|cc} 
\toprule
\multirow{2}{*}{} & \multicolumn{2}{c|}{Hot-diverse 11\%} & \multicolumn{2}{c|}{Hot-narrow 15\%} & \multicolumn{2}{c}{Rare-narrow 73\%} \\ 
\cmidrule(l){2-7}
                  & \multicolumn{1}{c}{Hum} & \multicolumn{1}{c|}{Uni} & \multicolumn{1}{c}{Hum} & \multicolumn{1}{c|}{Uni} & \multicolumn{1}{c}{Hum} & \multicolumn{1}{c}{Uni}  \\ 
\midrule
One2One            & 93.8& 90.4 & 80.0& 61.2 & 96.0& 61.4 \\ 
\midrule
PUSL w/o PT             & \textbf{98.0} & \textbf{93.9} & \textbf{93.7}& \textbf{80.1} & \textbf{96.1}& \textbf{75.6} \\
\bottomrule
\end{tabular}
\caption{Human evaluation of top-5 keyphrase generation on \ourdata.
Hum: Percentage of human-judged relevant keyphrases.
Uni: Percentage of keyphrases present in the label universe.
One2One performs poorly in Hot-narrow, an inherently narrow-label subset, where forced overgeneration leads to low relevance and universe coverage. 
}
\label{tab:gtfree}
\vspace{-5mm}
\end{table}


\paragraph{Toward More Faithful Evaluation: F1 vs BudgetAccuracy.}
In Test-Narrow scenarios with sparse ground truth, F1 scores can misleadingly favor ``lazy'' models (Section~\ref{sec:metrics}). For instance, in AmazonCat-Test-Narrow, GROOV achieves the highest F1@$\mathcal{O}$ (62.54) by generating fewer keyphrases (2.99 on average when asked for 10), coincidentally matching the sparse ground truth, thus inflating its F1 score. In contrast, B@2$\mu$ provides a more balanced evaluation, revealing GROOV's limitations with the lowest score (24.34) when required to generate more keyphrases.

\subsection{Early-Termination Bias}
\label{sec:one2seq}
We compare PUSL with state-of-the-art GROOV and a plain One2Seq model, all from the One2Seq family. PUSL uniquely generates a flexible range of keyphrases on demand, overcoming self-terminating limitations.

In Table~\ref{tab:gtdependent}, when tasked with generating $\mu$ or $2\mu$ keyphrases, PUSL consistently generates more unique keyphrases (\#K@$\mu$ and \#K@$2\mu$) across all datasets. In \ourdata, where the average keyphrase number is 3, PUSL and the post-trained model generates 3-4 keyphrases when asked for 6 ($2\mu$), while GROOV produces only 1. PUSL's superior quality is reflected in higher F1@$\mathcal{O}$ and BudgetAccuracy scores. For example, in EURLex4.3k's Test-Diverse subset, PUSL achieves a B@$\mu$ of 88.56, significantly outperforming GROOV (69.06) and One2Seq (77.05). This consistent performance across diverse datasets demonstrates PUSL's ability to overcome early-termination bias, effectively addressing a critical OXMC limitation when required keyphrase numbers exceed training data averages.

\subsection{Overgeneration Bias}
\label{sec:one2one}
In evaluations with missing ground truth (Table~\ref{tab:gtdependent}), PUSL outperforms One2One in the high-$\mu$ EURLex4.3k dataset but shows mixed performance in the low-$\mu$ \ourdata dataset, highlighting the need for ground truth-free evaluation.

We conduct a human evaluation\footnote{We sampled 100 data points for each spilt.} on \ourdata test set samples, assessing keyphrase relevance to input text. We also check if predicted keyphrases exist in a million-level label set (user query universe), as absence may indicate impractical rarity. We report the average scores in Table~\ref{tab:gtfree}.
The table shows that One2One scores significantly lower than PUSL in Hot-Narrow human evaluation, likely due to forced generation of irrelevant keyphrases when annotators provide similar keyphrases. For the user query universe evaluation, both models score lower in the Narrow sets, highlighting challenges in predicting diverse keyphrases.

A case study on the ``PS4 Shenmue I \& II First Limited Edition Sony PlayStation 4 Japan Import'' shows that PUSL correctly generated keyphrases like \emph{shenmue ps4''}, aligning well with actual user searches, while One2One, despite generating relevant keyphrases, also produced nonexistent ones like \emph{``shenmue iii ps4''}. Additional case studies and a scalable ground truth-free evaluation are provided in Appendix~\ref{sec:gtfree_appendix}.

\begin{table}[t]
\centering
\small
\setlength{\tabcolsep}{1pt} 
\begin{tabular}{lrrrrrrr}
\toprule
Model    & P@O   & P@$\mu$    & P@2$\mu$   & B@$\mu$    & B@2$\mu$   & \#k@$\mu$  & \#k@2$\mu$ \\
\midrule
\multicolumn{8}{l}{\textbf{Test-Narrow}} \\
GROOV    & \textbf{54.8} & \textbf{54.8}  & \textbf{54.8}  & 18.7  & 9.4   & 1.0  & 1.0  \\
One2Seq  & 44.7 & 24.6  & 24.6  & 25.5  & 12.8  & 1.9  & 1.9  \\
PUSL     & 45.7 & 35.2  & 17.6  & \textbf{32.4}  & \textbf{18.5}  & \textbf{2.5}  & \textbf{3.4}  \\
\midrule
\multicolumn{8}{l}{\textbf{Test-Diverse}} \\
GROOV    & \textbf{73.4} & \textbf{73.4}  & \textbf{73.4}  & 26.0  & 13.0  & 1.1  & 1.1  \\
One2Seq  & 68.8 & 41.4  & 41.4  & 41.4  & 20.7  & 1.9  & 1.9  \\
PUSL     & 63.1 & 72.5  & 36.3  & \textbf{55.5}  & \textbf{33.4}  & \textbf{2.5}  & \textbf{3.3}  \\
\bottomrule
\end{tabular}
\caption{Comparison of traditional and proposed metrics for OXMC models.}
\vspace{-5mm} 
\label{tab:additional_metrics}
\end{table}

\subsection{Evaluation Metrics: P@k vs P@O vs B@k}
We compare traditional Precision@k (P@k) against our proposed metrics on \ourdata (Table~\ref{tab:additional_metrics}). The results quantify how P@k can mislead evaluation with incomplete ground truth - GROOV achieves high P@k scores (54.8 and 73.4) by generating only one keyphrase (1.0--1.1) that often matches a ground truth label. In contrast, B@k reveals GROOV's under-generation through significantly lower scores (9.4 and 13.0).

PUSL generates increasingly more unique keyphrases as k increases (2.5 $\rightarrow$ 3.4 Test-Narrow, 2.5 $\rightarrow$ 3.3 Test-Diverse). While this prolific generation leads to lower P@k compared to GROOV at higher k values, PUSL achieves consistently higher B@k scores (18.5 and 33.4), indicating its predictions better reflect the true label distribution despite incomplete ground truth.

\subsection{Comparison of PUSL vs GPT-4}
In this section, we compare PUSL with zero-shot GPT-4, representing two different paradigms. PUSL is a small domain-specific fine-tuned model, while GPT-4 is a general-purpose LLM using zero-shot learning\footnote{See Appendix~\ref{sec:gpt4gen_prompt} for the full prompt.}. This comparison aims to highlight the importance of domain-specific fine-tuning in generating relevant and query-aligned keyphrases.

In the \ourdata, GPT-4 achieved perfect human-judgment alignment but generated only 16\% of keywords found in the actual user query universe. In contrast, PUSL produced 72\% of its predicted keywords within the same universe, demonstrating its effectiveness in generating not only relevant but precisely aligned keyphrases due to domain-specific fine-tuning. This aligns with recent findings on the effectiveness of large language models for keyphrase generation \cite{song2023chatgpt,song2023large}. Although PUSL can be adapted to fine-tune large models, the computational demands are substantial given the OXMC data scale. However, this comparison underlines the benefits of PUSL's fine-tuning approach in generating keyphrases that meet real-world user needs.

\subsection{Exhaustive Annotation Experiment}
\label{sec:exhaustive}
XMC datasets often involve multiple annotators independently selecting data to annotate, leading to potential duplication and missing labels due to self-selection bias. Our proposed PUSL simulates a single annotator exhaustively annotating each data point.  We compared this with a human performing the same task.  Table~\ref{tab:exhaustive} reports the average number of unique keyphrases for ground truth (GT, multiple annotators), Human (single exhaustive annotator), and PUSL. Results show that in Hot-Diverse, GT (with thousands of annotators) generates significantly more unique keyphrases than a single Human annotator, who typically experiences cognitive overload after 10 keyphrases, often leading to repetition. For Hot-Narrow, Human produces only 1.8 keyphrases on average, highlighting the challenge of this split.

\begin{table}[t]
\footnotesize
\centering

\begin{tabular}{@{}lrrr@{}}
\toprule
         & GT & Human & PUSL w/o PT \\ \midrule
Hot-Diverse & 83.0    &  6.7     &   3.3   \\
Hot-Narrow  &   1 &   1.8    &  3.1    \\
Rare-Narrow  &   1 &    3.3   &     3.0 \\ \bottomrule
\end{tabular}
\caption{Comparison of the average number of unique keyphrases for Ground Truth (GT), Human exhaustive annotation, and PUSL (requested to generate 5 keyphrases). PUSL demonstrates consistent performance across scenarios, while humans struggled with Hot-Narrow annotation.}
\vspace{-5mm} 
\label{tab:exhaustive}
\end{table}

\section{Discussion}
\label{sec:discussion}

\paragraph{Annotation Strategies and Self-Selection Bias} 
OXMC datasets are typically sourced from two main types of data: e-commerce platforms (e.g., amazoncat) and community-driven websites (e.g., Wikipedia articles, Stack Overflow questions). E-commerce data relies on independent user annotations, where each user labels a self-selected subset of items. Community platforms employ collaborative editing, where users iteratively build upon previous annotations. This collaborative process typically yields richer keyphrase sets per item compared to independent annotations, given the same number of annotators. However, OXMC dataset curation obscures this distinction by retaining only unique labels, masking the underlying annotation bias~\cite{marlin2012collaborative,chen2023bias}.

\paragraph{Feedback Loops and Previous Model Bias}
Missing labels in e-commerce data also stem from a self-reinforcing cycle between system exposure and user behavior. Search algorithms favor frequently-queried items, increasing their visibility. Enhanced visibility leads to more user interactions, which in turn strengthen these items' positions in search results. This creates a Matthew effect, where popular items accumulate annotations while less popular ones remain under-annotated. Traditional OXMC datasets only record the final item-label associations without the underlying interaction patterns, perpetuating these biases. Models trained on such biased data may further amplify the disparity through ``previous model bias''~\cite{liu2020general}, creating feedback loops that reinforce already-popular items. This can lead to ``echo chambers'' where popular items dominate the label space while niche items become increasingly invisible ~\cite{jiang2019degenerate,ge2020understanding}.

\section{Conclusion}
This paper addresses missing labels in extreme multi-label classification datasets, attributing the issue to self-selection bias. We recast OXMC as Positive Unlabeled Sequence Generation to foster prolific and accurate label prediction, mitigating model laziness caused by missing training labels. Additionally, we propose B@$k$ metrics for faithful model assessment given incomplete ground-truth labels. Our approach bridges the gap between training label sparsity and inference-time prediction requirements in real-world OXMC applications.



\newpage

\section*{Limitation} 
First, we did not conduct missing label analysis on the popular datasets, because the annotation frequency information is unavailable.
Second, our current formulation focuses on generating keyphrases for individual items and does not directly address the task of item-item recommendation, because the model's open-vocabulary generation nature may make up nonexistent items. 

\bibliography{main}

\clearpage

\appendix

\begin{table*}[!t]
\centering
\small
\begin{tabular}{@{}l|rr|rr|rr|rr|rr|r@{}}
\toprule
Dataset   & $|\mathcal{D}_{\text{train}}|$ & $\overline{L_{\text{train}}}$ & $|\mathcal{D}_{\text{dev}}|$ & $\overline{L_{\text{dev}}}$ & $|\mathcal{D}_{\text{test}}^{\text{n}}|$ & $\overline{L^{\text{n}}_{\text{test}}}$ & $|\mathcal{D}_{\text{test}}^{\text{d}}|$ & $\overline{L^{\text{d}}_{\text{test}}}$ &  $L$ \\ \midrule
AmazonCat &  1,064,452                              &   5.06        &                 25,000             &    5.07      &       10,000              &     5.5      &        4,313                              &    14.52       &  13,330   \\
Wikipedia-1M  &      2,157,967                          &    4.90       &   142,645                           &  3.95         &       103,438                            &   3.89       &               10,128                     &   15.37        &  960,106   \\
EURLex-4.3K  & 42,771 & 15.94 &  6,000 & 15.92 &    3,053                        &    14.04      &    176                               &     37.53           &    4,271           \\
\ourdata     &   422,617            &    3.09       &    25,000                          &    3.10       &    5,152                               &    3.15               &   1,897                                &     16.91      &   233,612  \\ \bottomrule
\end{tabular}
\caption{$|\mathcal{D}_{\text{train}}|$: number of training instances, $\overline{L_{\text{train}}}$: average number of labels per training instance, $|\mathcal{D}_{\text{dev}}|$: number of development instances, $\overline{L_{\text{dev}}}$: average number of labels per development instance, $|\mathcal{D}_{\text{test}}^{\text{n}}|$, $|\mathcal{D}_{\text{test}}^{\text{d}}|$: number of instances in the narrow and diverse test sets, respectively, $\overline{L^{\text{n}}_{\text{test}}}$,  $\overline{L^{\text{d}}_{\text{test}}}$: average number of labels per instance in the narrow and diverse test sets, respectively, $L$: number of unique labels in the whole corpus.}
\label{tab:dataset_stats}
\end{table*}


\section{Data Details}
\label{sec:datadetail}

Prior to \cite{gupta2021generalized}, XMC datasets were not split from an independent and identically distributed (IID) perspective. Instead, they were split in a way that ensured every label had at least one training point. \cite{simig-etal-2022-open} further extended XMC to an open-vocabulary setting where models may be expected to generate unseen labels in the form of multi-token keyphrases. The baselines in this study are all capable of open-vocabulary generalization. Additionally, to evaluate models in a top-$k$ setting under different keyphrase diversity, we merge and re-split the datasets using Algorithm~\ref{algo:dataset_split}.

We also create for each dataset, 2 classes of sub-datasets for reporting aggregated top-$k$ numbers --- narrow ($D^n_{test}$), and diverse ($D^d_{test}$) which are aggregated from the sub-datasets generated in line 11 of Algorithm \ref{algo:dataset_split}. In terms of notation: \begin{equation} D^p_{test} = \bigcup_{i}^j D^k_{test} \end{equation} where i and j denote the lower and upper bounds for the percentile limits for the number of key-phrases per row in the test dataset and $p \in \{n,d\}$. For rare ($D^n_{test}$), $i=0$ and $j=2*\mu$ and for diverse ($D^d_{test}$), $i=2*\mu+1$ and $j=100^{th}$ percentile. The data statistics is in Table~\ref{tab:dataset_stats}.






\begin{algorithm}[t]
\small
\caption{Dataset Split Procedure}
\begin{algorithmic}[1]
\Require $\mathcal{D}_{\text{train}}$, $\mathcal{D}_{\text{dev}}$, $\mathcal{D}_{\text{test}}$ \LComment{Merge and shuffle datasets}
\State $\mathcal{D} \gets \text{shuffle}(\mathcal{D}_{\text{train}} \cup \mathcal{D}_{\text{dev}} \cup \mathcal{D}_{\text{test}})$ 
\LComment{Group by text input and concatenate labels}
\State $\mathcal{D} \gets \text{GroupBy}(\mathcal{D}, x).\text{Aggregate}(\text{concatenate})$ 
\LComment{Uniformly split the dataset}
\State $[\mathcal{D}_{\text{train}}, \mathcal{D}_{\text{dev}}, \mathcal{D}_{\text{test}}] \gets \text{UniformSplit}(\mathcal{D})$ 
\LComment{Split based on label number}
\State $l \gets \min(\{|y| \, \forall y \in \mathcal{D}\})$
\State $r \gets \max(\{|y| \, \forall y \in \mathcal{D}\})$

\For{$k = l$ to $r$}
    \State $\mathcal{D}_{\text{train}}, \mathcal{D}_{\text{test}}^k \gets \text{SplitByNum}(\mathcal{D}_{\text{train}}, k)$ 
\EndFor

\State \Return $\mathcal{D}_{\text{train}}, \mathcal{D}_{\text{dev}}, \mathcal{D}_{\text{test}}, \{\mathcal{D}_{\text{test}}^k\}_{k=l}^{r}$
\end{algorithmic}
\label{algo:dataset_split}
\end{algorithm}

\begin{table}[t]
\footnotesize
\centering
\begin{tabular}{lrrrrrr}

\toprule
Model & \#K@3 & \#K@6 & B@3 & B@6 & F1@O \\
\midrule
PUSL                     & 2.46 & 3.29 & 40.32 & 23.32 & \textbf{35.21} \\
PUSL + raw               & 2.26 & 2.75 & 62.33 & 36.17 & 13.94 \\
PUSL + synt              & \textbf{2.82} & 4.06 & 77.33 & 50.67 & 17.18 \\
PUSL + raw + synt        & 2.75 & \textbf{4.21} & \textbf{78.67} & \textbf{53.50} & 18.19 \\
\bottomrule
\end{tabular}
\caption{Performance comparison of PUSL models with various post-training data.}
\label{tab:super_pusl_ablation}
\end{table}

\section{Implementation Details}
\label{appendix:hyper}
All experiments were conducted using 4 $\times$ Tesla V100-SXM2-32GB GPUs. The PUSL training times for the datasets were as follows:
\begin{itemize}
\setlength{\itemsep}{-0.5em}
    \item Eurlex4.3k: 3 hours
    \item Wiki: 1 day 20 hours
    \item AmazonCat: 2 days 18 hours
    \item \texttt{\ourdata}: 1 day 2 hours
\end{itemize}

We used the T5-base architecture for all compared models with identical hyperparameters. Following \cite{simig-etal-2022-open}, we set the learning rate to 2e-4, number of epochs to 10, and total batch size to 32. For post-training, we constructed a dataset of 50,000 samples, halved the initial learning rate with linear decay to 0, and trained for 3 epochs.
When evaluating top-$k$ models under F@$\mathcal{O}$, we truncate the predicted keyphrases to a maximum of 100 to avoid memory limitations on our machine.

\section{Ablation study on post training}
\label{appendix:ppusl}
\begin{table}[t]
\footnotesize
\centering
\begin{tabular}{lrrrrrr}
\toprule
Model & \#K@3 & \#K@6 & B@3 & B@6 & F1@O \\
\midrule
PUSL                     & 2.46 & 3.29 & 40.32 & 23.32 & 35.21 \\
\midrule
One2Seq                  & 2.10 & 2.10 & 54.33 & 27.17 & 10.99 \\
One2Seq + synt           & 1.77 & 1.77 & 48.67 & 24.33 & 9.31 \\
One2Seq + raw            & 1.87 & 1.87 & 50.67 & 25.33 & 10.18 \\
\bottomrule
\end{tabular}
\caption{Post-training effects on One2Seq.}
\label{tab:super_ablation_one2seq}
\end{table}

We conducted an ablation study to examine the individual contributions of two data sources, raw and synthetic, used during post-training. This experiment was conducted on a super-diverse test set from Ads-XMC, where the average number of keyphrases per instance is 89.89, compared to 16.91 in the original diverse set. The larger number of keyphrases per instance provides a more comprehensive ground truth, allowing for a more thorough evaluation of the impact of each data source on model performance.

The results, as shown in Table~\ref{tab:super_pusl_ablation}, indicate that both data sources contribute to the overall performance, but in distinct ways. The raw data from the original Train-Diverse set helps maintain stable performance on well-represented inputs, though it does not significantly increase the number of generated keyphrases. In contrast, the synthetic data, augmented through rejection sampling, enhances the model's diversity by increasing the number of keyphrases generated, thus improving performance across various metrics.

When both raw and synthetic data are combined, the model achieves stronger overall performance compared to using either data source individually. This complementary effect suggests that raw data provides stability for common inputs, while synthetic data improves diversity and coverage. It is worth noting that as the model generates more keyphrases, the F1@O score decreases significantly, reflecting a tendency to favor more prolific models. However, as the model becomes more prolific, the budget accuracy score increases steadily, providing a more faithful and reliable measure of model performance

To investigate the effects of post-training on the baselines, we conducted additional ablation study for One2Seq, using both raw data from the original Train-Diverse set and synthetic data generated through PUSL’s rejection sampling method. However, it is important to note that post-training is not applicable to One2One, as it inherently generates the requested number of keyphrases without flexibility for further augmentation.

As shown in Table~\ref{tab:super_ablation_one2seq}, simply adding augmented data (either raw or synthetic) to the baseline One2Seq model does not resolve the core challenges of OXMC tasks. In fact, post-training with both data sources resulted in decreased performance for One2Seq, as evidenced by lower metrics across the table. Notably, the number of unique keyphrases generated by One2Seq after post-training (\#K) was lower than its original output, indicating a higher repetition rate. We hypothesize that this drop in performance is due to the early-termination bias ingrained in One2Seq during its initial training. Since One2Seq uses an EOS token to conclude keyphrase generation, it tends to terminate prematurely after learning this behavior over a large dataset. During post-training, the smaller amount of augmented data appears insufficient for the model to unlearn this termination behavior, thus limiting its ability to generate more diverse keyphrases

\begin{figure}[t!]
  \centering
    \includegraphics[width=0.75\linewidth]{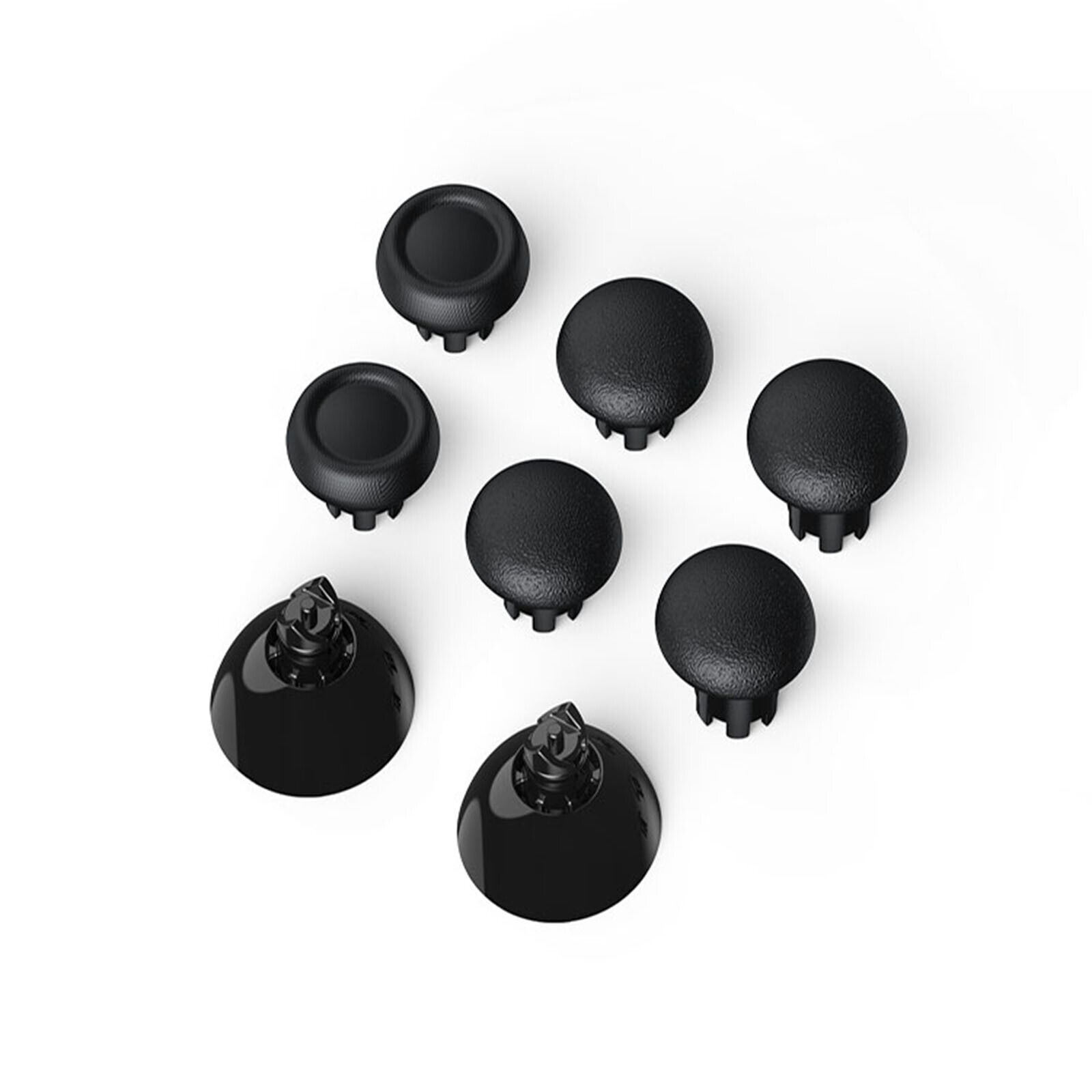}
\caption{Case study on One2One passing the Uni (user query universe) evaluation but failing on Human evaluation. The item title is ``Joystick Rocker Cap Buttons Cover Thumb Stick Grip Cap for PS5 DualSense Edge,'' but One2One predicts it as ``thumb grip.'' While the human annotator considers ``thumb grip'' as a made-up term, buyers may actually use it to search for joystick caps. This highlights the complexity of evaluating search term predictions, as informal or colloquial terms might be practically useful despite not being formally correct.
}
  \label{fig:appendix_case} 
  \vspace{-4mm}
\end{figure}

\section{LLM-as-a-Judge Evaluation}
\label{sec:gtfree_appendix}
While human evaluation, as conducted in Section~\ref{sec:one2one}, provides valuable insights, it is time-consuming and challenging to scale. Given the absence of comprehensive labels in the test data, we developed a ground truth-free evaluation method. We leveraged GPT-4 to simulate human evaluation, offering a scalable alternative. Our approach involves prompting GPT-4 with the input text and the keyphrases predicted by the models. GPT-4 then assesses the relevance of each keyphrase to the input, which we denote as binary score $G$. This GPT-4-based evaluator has been benchmarked against positive buyer judgment and achieved more than 90\% alignment, see \cite{ashirbad-etal-2024}. We define the final relevance score for each keyphrase as:
\begin{equation}
Rel = \frac{1}{N} \sum_{i=1}^{M} \sum_{j=1}^{K_i} (G_{ij} \land Uni_{ij})
\end{equation}
where $Rel$ is the overall relevance score, $M$ is the number of input texts, $K_i$ is the number of keyphrases for the $i$-th input text, $N = \sum_{i=1}^{M} K_i$ is the total number of keyphrases across all inputs, $G_{ij} \in \{0,1\}$ indicates whether GPT-4 judged the $j$-th keyphrase of the $i$-th input as relevant, and $Uni_{ij} \in \{0,1\}$ indicates whether the same keyphrase is present in the domain-specific universe of queries. The logical AND operation ($\land$) ensures that a keyphrase contributes to the relevance score only if it satisfies both general textual relevance and domain-specific considerations.

This methodology allows us to balance general textual relevance ($G$) with domain-specific considerations ($Uni$). While $G$ captures broader contextual relevance, it may not fully align with domain-specific user preferences, which are better represented by $Uni$. The results in Table~\ref{tab:gtfree_appendix} reveal that the GPT-4 evaluation tends to be more lenient than human assessment, often considering predicted keyphrases as relevant. However, the combined $Rel$ score provides a more faithful and scalable evaluation metric, effectively incorporating both GPT-4's textual relevance judgment and domain-specific considerations.

Figure~\ref{fig:appendix_case} illustrates a compelling case study that highlights the nuances and challenges in evaluating keyphrase predictions. The item in question, ``Joystick Rocker Cap Buttons Cover Thumb Stick Grip Cap for PS5 DualSense Edge,'' was predicted by the One2One model as ``thumb grip.'' While this prediction passed the Uni (user query universe) evaluation, indicating its presence in domain-specific queries, it failed the human evaluation. The human annotator likely considered ``thumb grip'' as a non-standard or colloquial term.  The combined $Rel$ score, therefore, provides a more comprehensive assessment that balances linguistic accuracy, domain specificity, and user behavior, offering a more nuanced and practically useful evaluation metric.



\section{Prompts for GPT-4}
\label{sec:prompts}

We use the GPT-4 API version gpt-4-0613 for all the experiments.
\paragraph{GPT-4 as a keyphrase relevance evaluator}
\label{sec:relevance_prompt}
Below is the prompt for GPT-4 as a keyphrase relevance evaluator, with templated values to be replaced by the item title, description, and the model generated keyphrases.

\noindent\texttt{\small{Given an item with title: \{\textrm{title}\} and description: \{\textrm{desc}\},
determine whether below keywords are relevant by giving yes or no one by one as answer: \\
1. \{\textrm{kw1}\};\\
2. \{\textrm{kw2}\};\\
3. \{\textrm{kw3}\};\\
4. \{\textrm{kw4}\};\\
5. \{\textrm{kw5}\}.\\
\#\#\#Response:
}}

\begin{table}[!t]
\scriptsize
\centering
\begin{tabular}{l|lll|lll|lll} 
\toprule
\multirow{2}{*}{} & \multicolumn{3}{c|}{Hot-diverse 11\%}                                & \multicolumn{3}{c|}{Hot-narrow 15\%}                                 & \multicolumn{3}{c}{Rare-narrow 73\%}                                \\ 


\cmidrule(l){2-10}
                  & \multicolumn{1}{l}{G} & \multicolumn{1}{l}{Uni} & \multicolumn{1}{l|}{Rel} & \multicolumn{1}{l}{G} & \multicolumn{1}{l}{Uni} & \multicolumn{1}{l|}{Rel} & \multicolumn{1}{l}{G} & \multicolumn{1}{l}{Uni} & \multicolumn{1}{l}{Rel}  \\ 
\midrule
One2One            & \textbf{92.2}         & {90.4}         & 84.0                     & 90.4                  & 61.2                  & 57.8                     & 93                    & 61.4                  & 57.8                     \\ 
\midrule
PUSL              & 90.9                  & \textbf{93.9}         & \textbf{85.9}            & \textbf{92.9}         & \textbf{80.1}         & \textbf{76.5}            & \textbf{93.7}         & \textbf{75.6}         & \textbf{72.3}            \\
\bottomrule
\end{tabular}
\caption{Scalable evaluation using GPT-4 as a simulated human evaluator, extending Table~\ref{tab:gtfree}. The combined relevance score ($Rel$) integrates GPT-4's assessment ($G$) with the domain-specific universe score ($Uni$) using a logical AND operation.}
\label{tab:gtfree_appendix}
\vspace{-4mm}
\end{table}


\paragraph{GPT-4 as a keyphrase generator}
\label{sec:gpt4gen_prompt}
Below is the prompt for GPT-4 as a keyphrase generator, with templated values to be replaced by the item title and description.

\noindent\texttt{\small{Given an item with title: \{\textrm{title}\} and description: \{\textrm{desc}\},
give 4 short cpc keywords for the item (all lowercase).
\\
\#\#\#Response:
}}






\section{Out-of-Time vs 
Independent-and-identically-distributed Evaluation}

Although all models are trained and analyzed under the independent and identically distributed (IID) assumption, they are expected to generalize to future data (out-of-time) that may contain seen/unseen input text, seen/unseen keyphrases, and even completely new entities, such as \emph{``iphone 20''}. Comparing our IID test split and out-of-time split, we found that the performance of PUSL indeed decreases in the future. For instance, in \ourdata, the F1 score drops from 32.9 on the Uniform Unseen split to 26.7 on the Future Unseen split (Table~\ref{table:wiki_seen_unseen}).

To further investigate this performance drop, we examined the out-of-time Wikipedia-1M dataset \cite{gupta2021generalized}. We discovered significant issues that led us to exclude it from our main experiment. Firstly, we found that 43\% of the input text in the Wikipedia-1M test set have occurred in the training data, but surprisingly, only 0\% of the ground truth labels remain the same (Table~\ref{table:wiki_seen_unseen}). This stark discrepancy between the input text overlap and the ground truth consistency violates the IID assumption and raises concerns about the validity of the dataset for evaluating model performance on future data. The F1 score on the Wikipedia-1M Future Seen split is only 1.3, compared to 33.4 on \ourdata, further highlighting the severity of this issue. In contrast, \ourdata OOT exhibits a more reasonable balance, with 53\% input text overlap and 45\% of the ground-truth labels remaining unchanged. This suggests that \ourdata OOT is a more suitable dataset for evaluating the generalization capabilities of XMC models on future data. Secondly, the lack of timestamp information in Wikipedia-1M and impedes further analysis of the models' performance on future data. Without accurate temporal information, it becomes challenging to assess how well the models adapt to evolving trends and new entities over time.

\begin{table}[!t]
\centering
\small
\begin{tabular}{@{}l|rrr|rrr@{}}
\toprule
              & \multicolumn{3}{c|}{Wikipedia-1M} & \multicolumn{3}{c}{\ourdata} \\ 
              \midrule
              & P       & R       & F       & P      & R      & F      \\ \midrule
Uniform Unseen      &  12.9       &  8.9       &   9.7      &  46.1      &   34.4     &    32.9    \\
Future Seen   &   1.2      &  1.7       &    1.3    &  37.1       &   44.7     &  33.4      \\ \midrule
Future Unseen &    11.3     &    7.3     &   8.2      &   21.9
     & 46.9       &   26.7     \\ \bottomrule
\end{tabular}
\caption{Evaluation on IID vs future. Seen/Unseen refers to the input text. In uniform split, we group by the text and concatenate the labels, so all text in the Uniform test split are unseen. We report the @4 and @5 scores for Wikipedia-1M and \ourdata respectively.
}
\label{table:wiki_seen_unseen}
\end{table}

Given these concerns, we decided not to include the Wikipedia-1M OOT dataset in our main experiment. While predicting the future might be an inherent challenge, the inconsistencies and limitations of the Wikipedia-1M OOT dataset make it unsuitable for drawing reliable conclusions about the models' generalization capabilities. Instead, we focus on evaluating the models using our IID test split and \ourdata OOT, which provide a more balanced and representative assessment of their performance.

\end{document}